\documentclass[english,iop, twocolumn]{revtex4}
\usepackage[T1]{fontenc}
\usepackage[latin9]{inputenc}
\usepackage{amsmath}
\usepackage{graphicx}
\usepackage{amssymb}
\usepackage{esint}

\makeatletter
\@ifundefined{textcolor}{}
{%
 \definecolor{BLACK}{gray}{0}
 \definecolor{WHITE}{gray}{1}
 \definecolor{RED}{rgb}{1,0,0}
 \definecolor{GREEN}{rgb}{0,1,0}
 \definecolor{BLUE}{rgb}{0,0,1}
 \definecolor{CYAN}{cmyk}{1,0,0,0}
 \definecolor{MAGENTA}{cmyk}{0,1,0,0}
 \definecolor{YELLOW}{cmyk}{0,0,1,0}
 }

\makeatother

\usepackage{babel}

\begin{document}

\title{Quantum Inductance and High Frequency Oscillators in Graphene Nanoribbons }

\author{Milan Begliarbekov, Stefan Strauf, Christopher P. Search}

\affiliation{Department of Physics \& Engineering Physics, Stevens Institute of
Technology, Hoboken NJ, USA}

\keywords{Graphene, Nanoribbon, Inductance, Capacitance, AC}
\begin{abstract}
Here we investigate high frequency AC transport through narrow graphene
nanoribbons with topgate potentials that form a localized quantum
dot. We show that as a consequence of the finite dwell time of an
electron inside the quantum dot (QD), the QD behaves like a classical
inductor at sufficiently high frequencies $\omega\gtrsim$50 GHz.
When the geometric capacitance of the topgate and the quantum capacitance
of the nanoribbon are accounted for, the admittance of the device
behaves like a classical serial RLC circuit with resonant frequencies
$\omega\sim100-900$ GHz and Q-factors greater than $10^{6}$. These
results indicate that graphene nanoribbons can serve as all-electronic
ultra-high frequency oscillators and filters thereby extending the
reach of high frequency electronics into new domains.
\end{abstract}
\maketitle

\section{Introduction}

The recent isolation of graphene \cite{novoseloc-science-2004}, a
two dimensional atomically thin crystal comprised of sp$^{2}$ hybridized
carbon atoms, sparked an unprecedented amount of research activity
aimed at understanding and exploiting the unusual properties of this
material. Early research efforts centered around understanding graphene's
fundamental properties, and their applications for electronic devices.
Graphene was shown to exhibit anomalous \cite{Zhang2005,Anomalous_QHE,Li2007}
and fractional \cite{Du2009,Bolotin2009} quantum hall effects, $\pi$-Berry
phase \cite{Zhang2005}, and be capable of ballistic \cite{Huard2007,Suspended_Graphene,Graphene_BN},
and coherent \cite{Miao2007} transport. Furthermore, its ultra high
mobilities, and the promise of realizing ballistic transport at elevated
temperatures attracted the attention of device physicists and engineers,
who soon showed that properties such as high room temperature mobilities
\cite{Wang2008}, excellent thermal conductivity \cite{Balandin2008},
and unusually high mechanical durability \cite{Lee2008}, render graphene
to be the ideal material for single molecule gas detectors \cite{Lu2009,Moradian2008},
high density capacitors \cite{Huang2007}, and most notably, ultra-high
frequency transistors, which were recently shown to be capable of
100 GHz operation \cite{Lin2010}, and predicted to be capable of
THz operating frequencies \cite{Emtsev2009,Ryzhii2009,Wright2009}.
Consequently, graphene is listed as one of the candidate materials
for post-silicon electronics on the International Technology Roadmap
for Semiconductors \cite{ITRS_MAP_2}. 

\begin{figure}
\includegraphics[scale=0.5]{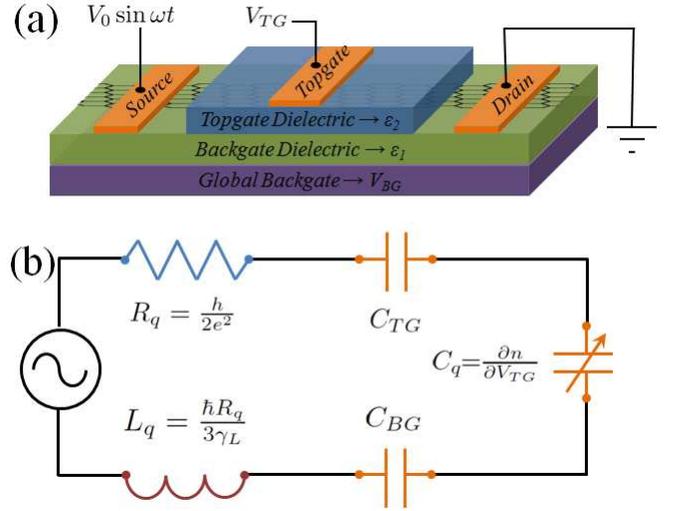}

\caption{(a) Schematic representation of the device and the biasing scheme,
(b) equivalent classical circuit model, in which $C_{q}$ is the quantum
capacitance of the graphene nanoribbon and $L_{q}$ is the quantum
mechanical inductance of the top gate defined quantum dot. $C_{BG}$
and $C_{TG}$ are the geometrical capacitances arising from the back
and top gate coupling respectively.}

\end{figure}

Although DC and low frequency transport in graphene nanostructures
is well understood \cite{Das_Sarma_REVIEW}, high frequency transport
in these devices has not yet been thoroughly investigated. Furthermore,
the sparse experimental measurements that characterize high frequency
graphene devices are limited by the maximum frequencies which can
be directly measured using commercial electronics. Consequently, the
cut-off frequency in these devices can only be extrapolated from its
$1/f$-gain plot \cite{Lin_Transistor_Nano,Lin2010a}, but not directly
measured. In this work, we examine AC transport in top-gated graphene
nanoribbons (GNRs) using the Green-Kubo linear response formalism.
We show that at sufficiently high frequencies, transport in GNRs is
analogous to a classical RLC circuit, where the inductive component,
which is quantum mechanical in origin, becomes dominant after a transition
frequency $\omega_{TR}$. Furthermore, if this inductive behavior
is coupled with the GNR's gate-tunable quantum capacitance, the resultant
circuit can be utilized as an all electronic ultra high frequency
oscillator. The ability to adjust quantum capacitance of the GNR oscillator
in-situ renders this device ideal for ultra high frequency all electronic
switching and measurement applications. Furthermore, we propose a
scheme in which GNR-based devices could be utilized to measure ultra
high frequency signals, thereby overcoming current measurement limitations.

\section{Theoretical Model}

A schematic representation of the device under consideration and a
biasing scheme are shown in Figure 1a. The device consists of a narrow
graphene nanoribbon of width $\lesssim50nm$, with Ohmic source and
drain contacts and capacitively coupled back and top gates, which
are electrically isolated from the GNR by two dielectrics with permettivities
$\varepsilon_{1}$ and $\varepsilon_{2}$. A thin metallic top gate
is utilized to provide a locally tunable barrier via the application
of an electrostatic bias $V_{TG}$, whereas a global bias may be applied
to the backgate $V_{BG}$, which is used to modulate the Fermi level
in the entire device. Furthermore, the top gate potential is used
to electrostatically define a quantum dot (QD). Similar devices are
routinely fabricated \cite{Huang2007,Stander2009}, and transport
in this structures has been studied in both DC \cite{Stander2009,Young2009}
and more recently low frequency AC \cite{Lin_Transistor_Nano} bias
regimes. In our discussion we assume that only a single level of the
QD is accessible to electrons in the source-drain bias window so that
the dot can be characterized by a single energy level $E_{0}$ and
line width $\gamma_{L}$.

In order to explore transport through this device in both DC and high
frequency AC regimes, we utilize the Green-Kubo formalism to calculate
the frequency dependent admittance $\Gamma_{D}\left(\omega,x\right)$
through the QD using

\begin{equation}
\Gamma_{D}\left(\omega,x\right)=\frac{1}{\omega L}\underset{-L/2}{\overset{L/2}{\int}}dy\underset{0}{\overset{\infty}{\int}}dte^{i\left(\omega+i\epsilon\right)t}\left\langle \hat{j}(x,t),\hat{j}(x,0)\right\rangle ,\end{equation}
where $L$ is the topgate length, $\hat{j}$ is the current operator,
and $\epsilon$ is a positive infinitesimal. The admittance for the
top gate defined quantum dot, which we denote as $\Gamma_{D}$ is
evaluated in Appendix I. Furthermore, in order to couple the high
frequency admittance of the QD to the GNR device, we utilize the method
introduced by Wang et al \cite{Wang_Inductance_PRB}: 

\begin{equation}
\frac{e^{2}}{C_{\mu}\left(\omega\right)}=\frac{e^{2}}{C_{0}}+\frac{\omega e^{2}}{i\Gamma_{D}\left(\omega,x\right)}+\frac{e^{2}}{C_{Q}},\end{equation}
where, $C_{0}$ is the total geometric capacitance, and $C_{Q}$ is
the quantum capacitance of the nanoribbon, which is proportional to
the density of states at the Fermi energy. The frequency dependent
electrochemical capacitance $C_{\mu}(\omega)$ is related to the admittance
of the entire device by $\Gamma(\omega)=-i\omega C_{\mu}(\omega)$.

\subsection{Quantum Inductance}

In order to interpret the physical meaning of the complex QD admittance
$\Gamma_{D}\left(\omega,x\right)\rightarrow\Gamma_{D}\left(\omega\right)$,
we express it in terms of its real and imaginary components $\Gamma_{D}\left(\omega\right)=\Re\Gamma_{D}\left(\omega\right)+i\Im\Gamma_{D}\left(\omega\right)$
(see Appendix I for the complete expression). Similar expressions
for the admittance are obtained in \cite{Wang_Inductance_PRB,Want_PRL}
using a nonequilibrium Green's function approach. We further introduce
a dimensionless parameter $\zeta\equiv2\left(E_{0}-\mu\right)/\gamma_{L}$
(see Appendix I) in order to characterize the width of the resonance,
where $\mu$ is the electrochemical potential of the source lead,
and the difference $E_{0}-\mu$ defines the transport window created
by the source-drain bias. In the limit $\omega\rightarrow0^{+}$,
we recover the DC Landauer conductivity $2e^{2}\left(h\left(1+\zeta^{2}\right)\right)^{-1}$.
A plot of both real $\Re\Gamma\left(\omega\right)$ and imaginary
$\Im\Gamma\left(\omega\right)$ components of the dynamic admittance
are shown in Appendix I. As can be seen in Fig. 7b, the sign of the
imaginary component of the admittance becomes positive after a critical
transition frequency $\omega_{TR}$, corresponding to a negative capacitance,
or, equivalently, to inductive behavior. It should be noted that in
quantum transport the sign convention is opposite to the one used
in electrical engineering. Namely, the sign of the frequency component
$\left(e^{-i\omega t}\right)$ is chosen such that $\Im\Gamma\left(\omega\right)>0$
corresponds to inductive behavior. Quantum inductance in various nanostructures
has been previously investigated both theoretically \cite{Yam_Nanotube_Inductance,Wang_Inductance_PRB,Fu_Inductance}
and verified experimentally \cite{Gabelli_Science}.

\begin{figure}
\includegraphics[scale=0.5]{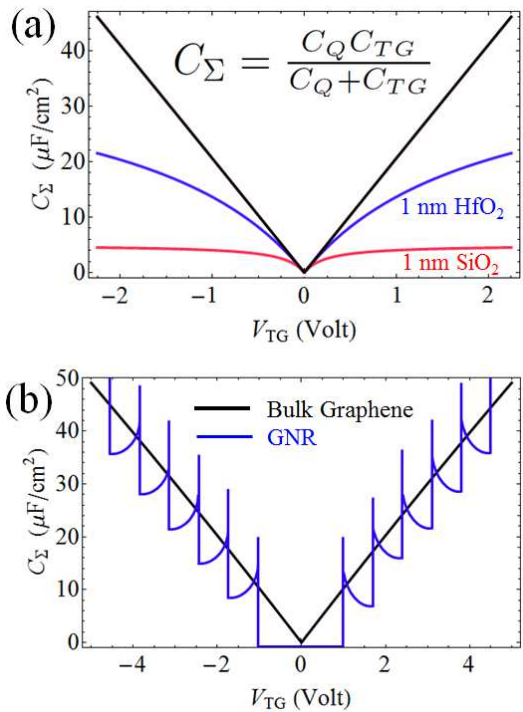}

\caption{(a) Total gate tunable capacitance $C_{\Sigma}$ of graphene using
suspended top gate (black), 1 nm SiO$_{2}$ (blue), and 1 nm HfO$_{2}$
(red) dielectrics; (b) a comparison between the quantum capacitance
of bulk graphene (black) and that of a 2.5 nm graphene nanoribbon
(purple) showing van Hove singularities on a suspended top-gate.}

\end{figure}

Phenomenologically, the inductive behavior of the QD can be understood
by viewing the QD and topgate as forming a parallel plate capacitor.
In classical parallel plate capacitors, charge accumulation on the
device results from the application of a voltage. However, even for
a quantum coherent capacitor there exists an intrinsic charge relaxation
resistance $R_{q}=h/2e^{2}=12.9\textrm{ k}\Omega$, in the limit of
a single transport channel \cite{Gabelli_Science,Nigg_PRL}. Consequently,
the charge accumulation time is $\tau_{RC}=(R_{q}C)^{-1}$. If the
driving voltage is an AC signal, the charge accumulation will follow
the voltage and will also reverse sign in due time . Ideally, in the
absence of resistance, there is a $\pi/2$ phase lag between the current
and the AC driving voltage. This is only true for low frequency AC
signals. For high frequencies $\omega\gg\tau_{RC}^{-1}$, the charge
buildup cannot follow the changes in the AC voltage. This situation
gives rise to an effective negative capacitance, or \textit{inductive}
behavior, since in this transport regime it appears as though the
voltage on the plates lags the current, as shown in the parametric
plot of the amplitude-phase diagram in Fig. 7d. It should be noted
that this effect arises from the presence of a quantum mechanical
charge relaxation resistance, $R_{q}$, and is therefore quantum mechanical
in origin. Consequently, quantum mechanical devices exhibit inductive
behavior at sufficiently high frequencies even though no geometric
inductor is present in the circuit. 

\begin{figure}
\includegraphics[scale=0.45]{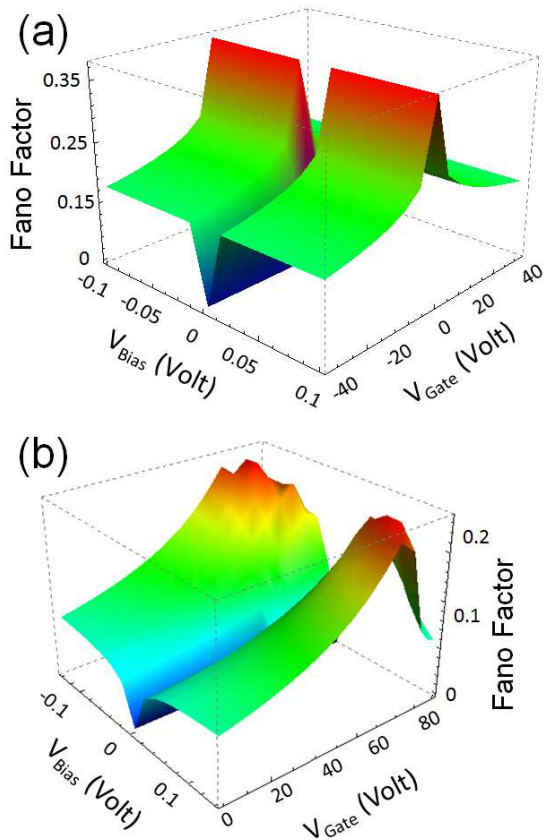}

\caption{Gate and bias voltage dependence of the Fano Factor of a GNR at $\omega=800$
MHz and (a) $W/L=25$ and (b) $W/L$ = 5, with $L=200$nm. These parameters
were chosen to match the experiments in Ref. \cite{Graphene_Shot,Danneau_Full_Shot_Noise}.}

\end{figure}

It was further pointed out \cite{Wang_Inductance_PRB}, that a second
relevant time scale in high frequency transport of QDs is introduced
by the carrier dwell time $\tau_{D}$ of the QD, which is related
to the line width function of the dot barrier $\gamma_{L}=4\hbar/\tau_{D}$
. Smaller $\gamma_{L}$ serves the purpose of increasing the charge
carrier dwell time inside the QD region and consequently lowering
the transition frequency $\omega_{TR}$ at which quantum inductance
appears. For sufficiently small $\gamma_{L}$ (i.e., large dwell times),
the transport is always inductive, even at low frequencies; however,
these regimes have not yet been realized experimentally due to the
difficulty that arises in fabricating strongly coupled topgate leads.
Utilizing the above considerations, Wang et al., \cite{Wang_Inductance_PRB}
define the quantum inductance as $L_{q}=R_{q}\tau_{d}/12$. In essence,
the finite line width forces a charge carrier to remain inside the
dot for at least time $\tau_{D}$ implying that the charge carriers
can only follow changes in the voltage for frequencies $\omega\ll1/\tau_{D}$.
At higher frequencies this effective trapping of the charges inside
the QD for $\tau_{D}$ gives rise to such large phase delays between
the current and voltage that the capacitance appears negative. In
our system, this dwell time effect is larger than the charge relaxation
time, $\tau_{D}>\tau_{RC}$ so that the inductive behavior is attributable
to the trapping of charge carriers in the QD.

\begin{figure*}
\includegraphics[scale=0.65]{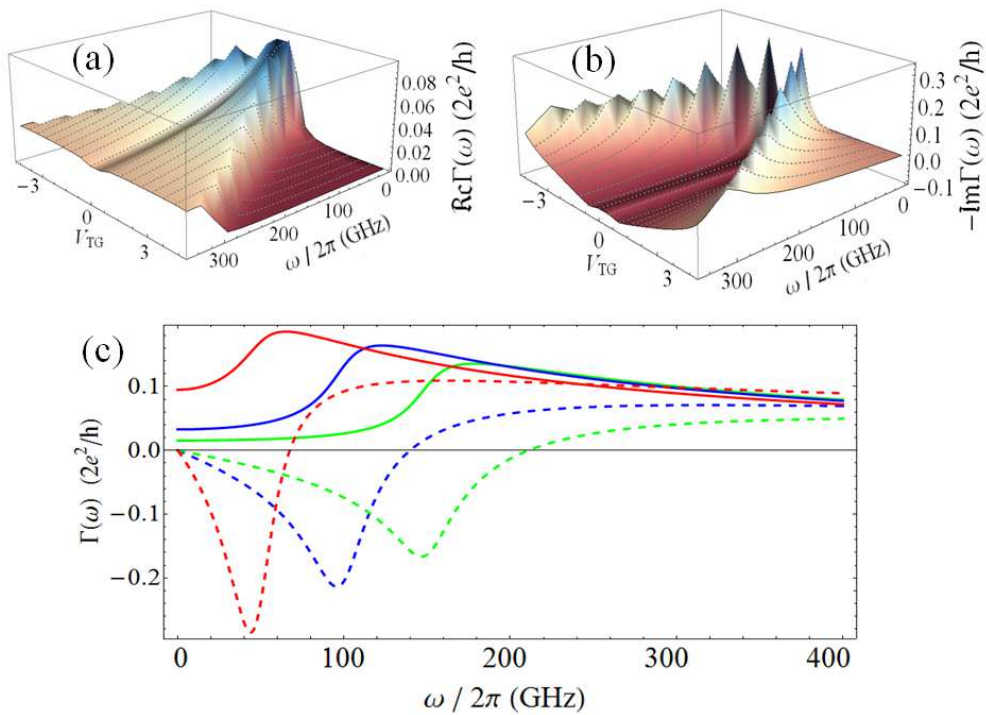}

\caption{(a) real and (b) minus the imaginary components of the dynamic admittance
of a 2.5 nm nanoribbon structure as shown in Figure 1a as a function
of $V_{TG}$(Volt) and frequency, with $E_{0}-\mu=50$mV, L = 200nm;
(c) real (solid) and imaginary (dashed) components of the dynamic
admittance for different values of the line width function $\gamma_{L}=0.1$
meV (red), $\gamma_{L}=0.3$ meV (blue), $\gamma_{L}=0.5$ meV (green),
using $V_{TG}=0.5$V, using a 300 nm SiO$_{2}$ backgate and a 10
nm HfO$_{2}$ top gate at T = 5 K.}

\end{figure*}

\subsection{Quantum Capacitance}

Unlike quantum inductance which is does not depend on graphene's particular
density of states, quantum capacitance, introduced by S. Luryi \cite{Luryi_Q_Cap},
depends on the underlying band structure of the material. The quantum
capacitance $C_{Q}$ describes the movement of the conduction band
as a function of the applied gate bias: $C_{Q}\equiv\left.e^{2}\frac{dn}{dE}\right|_{E=E_{F}}$.
Unlike in most conventional semiconductors, in graphene, the quantum
capacitance is an important parameter since in the low bias regime,
monolayer graphene exhibits a linear gate tunable dispersion \cite{Das_Sarma_REVIEW,Chen2008,Fang_APL_Q_Cap}.
In the case of narrow constrictions, when the GNR width approaches
the Fermi wavelength of the electrons, a bandgap opens \cite{Kim_GNR_Gap}
and $C_{Q}$ (as well as the density of states) exhibit van Hove singularities,
as shown in Fig. 2b \cite{Fang_APL_Q_Cap}. Consequently the quantum
capacitance of a graphene nanoribbon (per unit width) can also be
tuned by varying the gate bias according to

\begin{equation}
C_{Q}\left(V_{TG}\right)\cong\frac{4e^{2}}{\pi\hbar v_{F}}\underset{n}{\sum}\frac{\eta}{\sqrt{\eta^{2}-x_{n}^{2}}}\Theta\left(\eta-x_{n}\right),\end{equation}
where $\eta=E_{F}/k_{B}T$ , $x_{n}=E_{n}/k_{B}T$ is the energy of
the $\textrm{\ensuremath{n^{th}}}$ subband, and $\Theta\left(\eta-x_{n}\right)$
is the Heaviside step function. The expression for the quantum capacitance
of graphene stated in Eq. 3 is the same as the one obtained by other
authors \cite{Fang_APL_Q_Cap}, and it's derivation is outlined in
Appendix II. We can combine the geometric capacitances of the gates
and the quantum capacitance of the GNR to form the total capacitance

\[
\frac{1}{C_{\Sigma}}=\frac{1}{C_{Geometric}}+\frac{1}{C_{Q}}\]
 Since current graphene-based devices are fabricated using relatively
thick backgate dielectrics (typically on the orders of hundreds of
nanometers), $C_{BG}$, consequently, has no contribution to $C_{Geometric}$.
We therefore only need to take into account the contribution from
the topgate dielectric and can ignore the capacitive contribution
from the backgate and the capacitive coupling to the contacts, which
cannot be tuned by a gate bias. Although the capacitive coupling to
the backgate and contacts changes the value of the geometric capacitance,
which in turn decreases the voltage range over which $C_{\Sigma}$
can be modulated (a similar effect is achieved in Fig. 2a by changing
the dielectric constant of the top gate), gate modulation of the total
capacitance has been experimentally observed \cite{Nature_Q_Cap,Q_Cap2}.
The total capacitance $C_{\Sigma}$ is then determined by, $C_{\Sigma}=\frac{C_{Q}C_{TG}}{C_{Q}+C_{TG}}$.
The contribution of the quantum capacitance dominates in the limit
of small quantum capacitance $C_{Q}$, i.e., in devices with thin
gate dielectrics since it is in series with the geometric capacitance,
and for capacitors in series, the smaller capacitance dominates \cite{Chen2008,Fang_APL_Q_Cap,Nature_Q_Cap,Nano_Lett_Q_Cap}.
In typical GNR devices the oxide capacitance is on the order of $C_{oxide}\approx115aF/\mu m^{2}$
\cite{Huard2007} and the trap capacitance, $C_{tr}$, which arises
from the coupling to the leads is on the order of $C_{tr}\approx10fF/\mu m^{2}$.
while the quantum capacitance can be tuned in the range $5-35fF/\mu m^{2}$
\cite{Chen2008}. Figure 2a shows the total capacitance as a function
of several typical dielectric materials. The quantum capacitance of
a 2.5 nm GNR is shown in Fig 2b. While optical properties and carrier
transport in GNR\textquoteright{}s are in general affected by the
particular graphene edge state such as armchair, zigzag, or mixed
edges \cite{Micro_Raman_OUR}, for simplicity we consider GNRs with
pure armchair boundaries \cite{GNR_Edge_Theory}.

\subsection{Fano Factor}

In order to compare our model with previous experimental work, we
note that the real part of Eq. 1 is proportional to the frequency
dependent shot noise \cite{Djuric_Spin,Buttiker1993}. Consequently,
our model can be utilized to calculate the frequency dependent Fano
factor of a graphene nanoribbon. The dependence of the Fano factor
of a GNR on gate and source-drain bias at $\omega=800$ MHz is plotted
in Figure 3. The device parameters, such as the dielectric thickness,
GNR aspect ratio ($W/L=25,$ with $L=200nm$) and operating frequency
were chosen to match existing experimental results \cite{Graphene_Shot,Danneau_Full_Shot_Noise}.
The resultant dependence of the Fano factor is in excellent agreement
with the measured values. Experimentally, the Fano factor peaks at
a value $\mathcal{F}=0.34$, while in our model it peaks at $\mathcal{F}=0.46.$
Furthermore, the change in slope of the Fano factor occurs at $\mathcal{F}\thickapprox0.15$
in both the experimental data and in our calculation. Although, in
our calculation the Fano plot is symmetric about $V_{bg}=0$, this
is not the case in the experimental data. This discrepancy is caused
by the shift of the Dirac point in the presence of charged impurities
\cite{Tan2007,Begliarbekov_Oscillations}. The slight discrepancy
between Fig. 3a and the measured value at zero gate bias and the theoretical
prediction most likely stems from a finite density of states at the
Dirac point, which is a feature measured in numerous experiments \cite{Q_Cap_Ensslin}.
Although the above simulations model a relatively wide GNR, since
the real part of the frequency dependent admittance, is proportional
to the frequency dependent shot noise, the agreement between the experimental
data and our model point to its validity. However, further experiments
are needed for smaller GNRs at higher frequency to fully assess the
theoretical predictions.

\section{Results \& Discussion}

We now turn to the discussion of our main result, the dynamic admittance
of the top gated GNR. The real and imaginary components of $\Gamma(\omega)$
are plotted in Figure 4. Figure 4b, shows the imaginary component
of the admittance as a function of frequency and the top gate bias
(the sign of the imaginary admittance has been inverted for clarity)
where one can see that $-\Im\Gamma(\omega)$ becomes negative, corresponding
to an inductive behavior of the device. The spikes in $\Im\Gamma\left(\omega\right)$
correspond to resonances between the electronic states inside the
QD and the van Hove singularities of the GNR. Off-resonant cases are
plotted if Fig. 4c. In general, the width of the GNR changes the energy
spacing of the van Hove singularities. For a GNR with W=2.5nm, the
spacing is \textasciitilde{}1V (see Fig. 2b), while for 5 nm GNR the
energy spacing decreases to \textasciitilde{}0.4V \cite{Fang_APL_Q_Cap}.
Consequently, in order to achieve the desired resonance source-drain
voltages below the energetic spacing of the van Hove singularities
should be selected, in our case $V_{sd}=50mV$. Larger source drain
volrages (or equivalently wider GNRs) would result in several van
Hove peaks being inside the transport window. If a sufficiently small
number of resonance levels (2 or 3) are present in the transport window,
the sharp resonances in the conductance are no longer visible, however
the general features of the complex admittance are still present.
For a larger number of resonant levels, the transport becomes diffusive,
and the effect of quantum inductance is no longer observed.

In the limit $\omega\rightarrow0^{+}$ and large $\gamma_{L}$ the
real part of the frequency dependent impedance $Z\left(\omega\right)=\left(\Re\Gamma\right)^{-1}\rightarrow h/2e^{2}$
approaches the charge relaxation resistance $R_{q}$ as shown in Figure
5. For small values of $\gamma_{L}$, $Z\left(\omega\right)$ diverges
since low values of the line width function correspond to large values
of the carrier dwell times $\tau_{D}$, for which transport through
the QD becomes blocked. 

\begin{figure}
\includegraphics[scale=0.35]{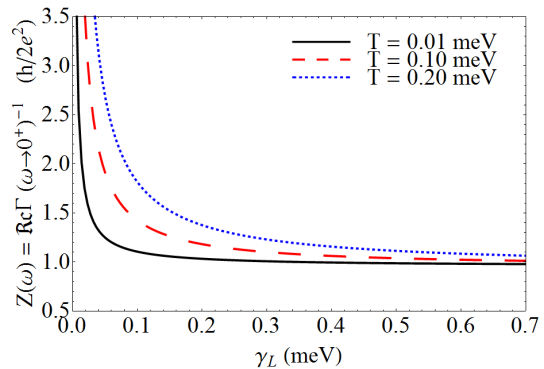}

\caption{Frequency dependent impedance $Z\left(\omega\right)$ in the limit
$\omega\rightarrow0^{+}$ at various temperatures for the same device
parameters as in Fig 4 ($V_{TG}=50$ mV), and the same parameters
as in Fig. 4.}

\end{figure}

For frequencies greater than $\omega_{TR}$ where $\Im\Gamma\left(\omega\right)>0$,
the transport through the GNR crosses over from being purely capacitive
to RLC behavior and the admittance of the device can modeled by a
classical series RLC circuit as

\[
\Gamma(\omega)=\frac{-i\omega C}{1-\omega^{2}L_{q}C-i\omega CR_{q}}.\]
The comparison of the classical RLC model to the quantum model is
shown in Fig. 6. In the above model, the circuit inductance of the
GNR is given by $L_{q}=h^{2}/12\pi e^{2}\gamma_{L}$ and $C=C_{\mu}(0)$.
In accordance with the model proposed by Wang et al. for a generic
quantum dot \cite{Wang_Inductance_PRB}, the transition frequency
$\omega_{TR}$ in our device occurs below the resonance frequency
$\omega_{0}$. The transition frequency is determined by the carrier
dwell time $\tau_{D}$ (or equivalently $\gamma_{L})$ and ranges
from 50 - 200 GHz for the values of $\gamma_{L}$ shown in Fig. 4,
which corresponds to inductances of 40 - 200 nH. Furthermore, the
resonance frequency $\omega_{0}$ of the resultant RLC circuit is
given by $\omega_{0}=\left(L_{q}C\right)^{-1/2}$. Using the above
inductance values, for typical quantum dot sizes of 100 nm$^{2}$
and 25 nm$^{2}$ fabricated on 300 nm SiO$_{2}$ with a capacitance
of 11.5 nF / cm$^{2}$ and 30 $\mu$F/cm$^{2}$ quantum capacitance
of graphene, we predict resonant frequencies of this oscillator on
the order of 100 - 900 GHz. This gives rise to Q-factors $Q=\frac{\omega_{0}}{\Delta\omega}=\frac{1}{R}\sqrt{\frac{L}{C}}$
on the order of Q = $1\times10^{6}$ - $8\times10^{6}$ ($\triangle\omega$
is the width of the resonance). This is a remarkable number for an
all electronic circuit. To compare this result to conventional 2DEG
structures, such as AlGaAs heterojunctions, we note that the dwell
time may be approximated as $\tau_{D}\sim4L/v_{F}$ \cite{Wang_Inductance_PRB},
where $L$ is the device length and $v_{F}$ is the Fermi velocity.
Since $v_{F}$ is AlGaAs 2DEGs is typically on the order of $\sim3\times10^{7}m/s$,
the corresponding Quantum inductance and Q factor would be an order
of magnitude lower. It should be noted that the above results were
obtained assuming prestine nanoribbons, and effects that would lead
to dephasing or the degradation of the ballistic mean free path, such
as the morphology \cite{Cones} of the device and potential fluctuations
have not been considered and are beyond the scope of this study. In
general, spatial potential fluctuations in graphene are on the order
of tens of meV \cite{Suspended_Graphene}, while the typical values
of $\gamma_{L}$ are on the order of $\mu$eV. The average size of
the potential fluctuations was measured to be \textasciitilde{}30
nm \cite{e_h_puddles}. Consequently, if the topgate length is smaller
than the size of the potential fluctuations (such as the case with
Si nanowires and carbon nanotube electrodes), the size of the potential
would be constant and would not affect the result. Furthermore $\gamma_{L}=10\mu$eV
corresponds to a dwell time of 260ps. Therefore, temporal potential
fluctuations that are the same order of magnitude as the dwell time
could potentially lead to a loss of coherence. However, if the GNR
is fabricated on Boron Nitride dielectrics \cite{Graphene_BN} or
is suspended \cite{Suspended_Graphene} the undesired effects of potential
fluctuations are largely reduced. 

\begin{figure}
\includegraphics[scale=0.5]{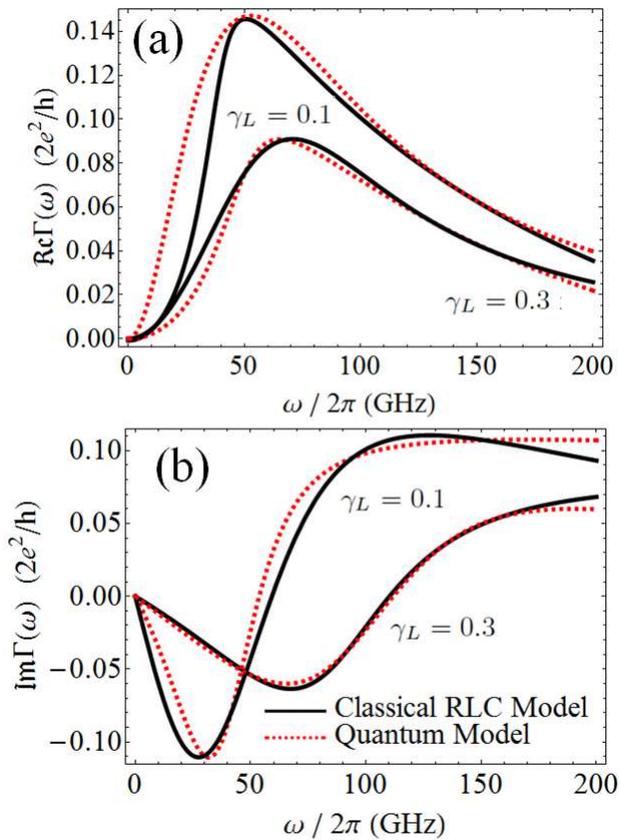}

\caption{A Comparison of (a) real and (b) imaginary admittance components of
the classical RLC model (solid black line) with the results of our
simulation (dotted red line) $\gamma_{L}=0.1$ meV and $\gamma_{L}=0.3$
meV for a 200 x 200 nm quantum dot (which is easily fabricated using
modern technology), and the same parameters as in Fig. 4.}

\end{figure}

The large Q-factor and in-situ tunability of the quantum capacitance
afforded by shifting the gate bias makes this device ideal for ultra-high
frequency electronic applications. Commercially available high frequency
electronic circuits are not capable of operating frequencies higher
than 40 GHz. However, the proposed device architecture, would extend
the operating frequency of all-electronic devices into the terahertz
regime. For example, if a high frequency AC signal is applied to the
source-drain contacts, a subsequent sweep of the top gate voltage
would drive the oscillator in and out of resonance, allowing for an
all electronic measurement of this frequency via the device gain.
Consequently, a measurement of the on and off resonant gain in the
device yields an all electronic measurement of the frequency of the
signal. Since $\omega_{0}$ is on the order of 100+ GHz, this GNR-RLC
circuit could be used to measure frequencies which are currently unattainable
by standard instrumentation. This, coupled with graphene's intrinsically
ultra-high carrier mobilities, render it an ideal material for all-electronic
THz devices.

\section{Conclusion}

In conclusion, we have applied the Green-Kubo formalism to model high
frequency transport through dual gated graphene nanoribbons. We showed
that above a sufficiently high frequency $\omega_{TR}$ determined
by the dwell time of charge carriers in the gate-defined quantum dot,
the behavior of the GNR device is analogous to a classical RLC oscillator
with a very high Q-factor. The inductive behavior arises from the
negative capacitance of the QD which occurs when charge carriers become
trapped inside the dot for times $\tau_{D}$ and cannot follow the
driving voltage. This leads to a phase lag greater than $\pi$ and
thus results in a negative capacitance. Coupling the inductive behavior
of a quantum dot to the gate tunable quantum capacitance in graphene,
gives rise to an in-situ tunable ultra high frequency oscillator and
filter thereby extending the reach of high frequency electronics into
the THz regime.
\begin{acknowledgments}
M.B. acknowledges partial financial support for this work that was
provided by the NSF GK-12 Grant No. NSF DGE-0742462 and S.S. acknowledges
support from the Air Force Office for Scientific Research (award no
FA9550-08-1-0134). C.P.S acknowledges support from NSF Grant. No.
0757933.
\end{acknowledgments}

\section*{Appendix I}

Here we derive the complex admittance for a QD with a single resonant
level. We emphasize that these results are general for any QD and
are not specific to graphene. Following the procedure in \cite{Fu_Inductance},
the expression in Eq. 1 for the admittance may be expressed in terms
of transmission $T\left(E\right)$ and reflection $R\left(E\right)$
amplitudes as

\[
\Gamma_{D}\left(\omega\right)=\frac{2e^{2}}{h}\frac{i}{2\pi\omega}\overset{\infty}{\underset{-\infty}{\int}}dE_{1}\overset{\infty}{\underset{-\infty}{\int}}dE_{2}\frac{f\left(E_{1}\right)-f\left(E_{2}\right)}{\omega+E_{1}-E_{2}+i\epsilon}\times\]

\begin{equation}
\Re\left\{ T\left(E_{1}\right)T^{*}\left(E_{2}\right)+1-R\left(E_{1}\right)R^{*}\left(E_{2}\right)\right\} ,\end{equation}
where $f$ is the Fermi function. The transmission and reflection
amplitudes through the barrier are given by 

\[
T\left(E\right)=\frac{i\gamma_{L}/2}{E-E_{0}+i\gamma_{L}/2}\]

\begin{equation}
R\left(E\right)=\frac{E-E_{0}}{E-E_{0}+i\gamma_{L}/2},\end{equation}
where $\gamma_{L}$ is a linewidth function representing the coupling
of the lead to the QD and characterize the energy width of the resonance
\cite{Fu_Inductance}. The linewidth function of the lead is related
to the carrier dwell time, $\tau_{D}$, inside the QD according to 

\[
\gamma_{L}=\frac{4\hbar}{\tau_{D}}.\]
Substituting Eq. 5 into Eq. 4 and carrying out the integrations, we
obtain the following result for the dynamic impedance of the QD $\Gamma\left(\omega\right)=\Re\Gamma\left(\omega\right)+i\Im\Gamma\left(\omega\right)$:

\[
\Re\Gamma_{D}\left(\omega\right)=\frac{2e^{2}}{h}\frac{\gamma_{L}}{4\hbar\omega}\times\]

\[
\left[\arctan\frac{-\zeta\gamma_{L}/2+\hbar\omega}{\gamma_{L}/2}-\arctan\frac{-\zeta\gamma_{L}/2-\hbar\omega}{\gamma_{L}/2}\right]\]

\[
\Im\Gamma_{D}\left(\omega\right)=\frac{2e^{2}}{h}\frac{\gamma_{L}}{8\hbar\omega}\times\]

\begin{equation}
\ln\left[\frac{\left[\left(\zeta\gamma_{L}/2+\hbar\omega\right)^{2}+\left(\gamma_{L}/2\right)^{2}\right]\left[\left(\zeta\gamma_{L}/2-\hbar\omega\right)^{2}+\left(\gamma_{L}/2\right)^{2}\right]}{\left[\left(\zeta\gamma_{L}/2\right)^{2}+\left(\gamma_{L}/2\right)^{2}\right]^{2}}\right],\end{equation}
The real and imaginary components of $\Gamma_{D}\left(\omega\right)$
are plotted in Figure 7 as a function of the dimensionless parameter
$\zeta$ (see main text for discussion).

\begin{figure}
\includegraphics[scale=0.36]{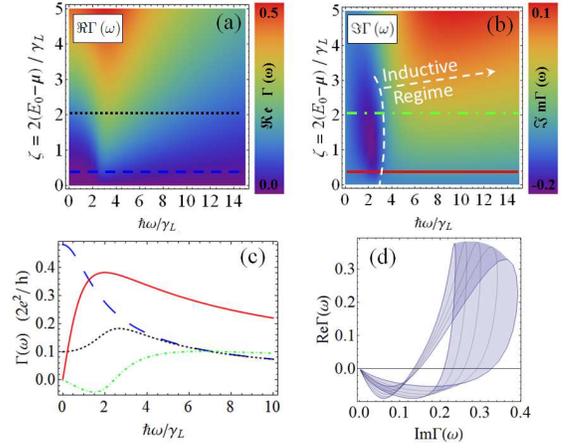}

\caption{(a) Real and (b) Imaginary components of the quantum dot admittance
(which does not include the contribution of the GNR) $\Gamma_{D}\left(\omega\right)$;
(c) several cuts through (a) at $\zeta=2$ (dotted black line), $\zeta=0.2$
(dashed blue line) and (b) at $\zeta=2$ (dot dashed green line) and
$\zeta=0.2$ (solid red line); $\Im\Gamma_{D}\left(\omega\right)>0$
corresponds to an inductive behavior; (d) parametric plot of real
and imaginary components as a function of $\omega$ for the same values
of $\zeta$ as in (c);}

\end{figure}

\section*{Appendix II}

In this section we provide a brief derivation of the quantum capacitance,
$C_{Q}\equiv\left.e^{2}\frac{dn}{dE}\right|_{E=E_{F}}$. To do so
we require an energy dependent expression for the density of states.
This is computed in the Landauer-B$\ddot{\textrm{u}}$ttiker formalism.
In the low bias regime the ballistic conductance $G^{(B)}$ is given
by

\[
G^{(B)}=\frac{2e^{2}}{h}\underset{-\infty}{\overset{\infty}{\int}}dET\left(E\right)M\left(E\right)\frac{-\partial f_{0}}{\partial E},\]

where $f_{0}$ is the Fermi function. In the ballistic limit $T\left(E\right)\rightarrow1$.
Furthermore, the mode function $M\left(E\right)$ is given by

\[
M\left(E\right)=\frac{\pi\hbar}{v}\underset{k}{\sum}\delta\left(E-E\left(\overrightarrow{k}\right)\right)\left|v_{g}\left(\overrightarrow{k}\right)\right|,\]

where $v_{g}\left(\overrightarrow{k}\right)$ is the group velocity
given by $v_{g}\left(\overrightarrow{k}\right)=\frac{1}{\hbar}\frac{\partial E}{\partial k_{x}}.$
Using this, we can express the mode function as

\[
M\left(E\right)=\frac{WL}{4\pi^{2}}\overset{\pi}{\underset{-\pi}{\int}}d\theta\overset{\infty}{\underset{0}{\int}}kdk\delta\left(E-E\left(\overrightarrow{k}\right)\right)\left|\frac{\hbar v_{F}k_{x}}{k}\right|,\]

which becomes,

\[
M\left(E\right)=\frac{2W}{\pi\hbar}\frac{\left|E\right|}{v_{F}},\]

where the factor 2 is introduced to account for spin degeneracy. Inserting
the above expression for $M(E)$ into $G^{(B)}$, and making use of
the fact that$\frac{-\partial f_{0}}{\partial E}=\frac{e^{\left(E-E_{F}\right)/k_{B}T}}{\left(1+e^{\left(E-E_{F}\right)/k_{B}T}\right)^{2}}$,
we obtain

\[
G^{(B)}=\frac{2e^{2}}{h}\frac{2W}{\hbar\pi v_{F}}\left\{ \overset{\infty}{\underset{0}{\int}}dEE\frac{e^{\left(-E-E_{F}\right)/k_{B}T}}{\left(1+e^{\left(-E-E_{F}\right)/k_{B}T}\right)^{2}}\right.\]

\[
\left.+\overset{\infty}{\underset{0}{\int}}dEE\frac{e^{\left(E-E_{F}\right)/k_{B}T}}{\left(1+e^{\left(E-E_{F}\right)/k_{B}T}\right)^{2}}\right\} \]

Integrating the above expression by parts, and doing some algebra,
we obtain

\[
G^{(B)}=\frac{2e^{2}}{h}\frac{2Wk_{B}T}{\hbar\pi v_{F}}\Gamma\left(1\right)\left\{ \mathcal{F}_{0}\left(-\eta_{F}\right)+\mathcal{F}_{0}\left(\eta_{F}\right)\right\} ,\]

where $\Gamma\left(x\right)$ is the Gamma function, $\mathcal{F}_{s}\left(\eta\right)$
is the Fermi-Dirac Integral given by $\mathcal{F}_{s}\left(\eta\right)=\frac{1}{\Gamma\left(s+1\right)}\overset{\infty}{\underset{0}{\int}}\frac{\varepsilon^{s}d\varepsilon}{1+e^{\varepsilon-\eta}}$,
and $\eta_{F}=\frac{E_{F}}{k_{B}T}$. Taking the low temperature limit
we have $\frac{-\partial f_{0}}{\partial E}\rightarrow\delta\left(E-E_{F}\right)$,
in which case the expression for the ballistic conductivity reduces
to

\[
G_{T=0}^{(B)}=\frac{2e^{2}}{h}\frac{2W\left|E_{F}\right|}{\pi\hbar v_{F}}.\]

These results can now be used to calculate the quantum capacitance
of a graphene strip defined as $C_{Q}\equiv\left.\frac{dn}{dE}\right|_{E=E_{F}}.$
Following the same procedure as above, the carrier densities are given
by

\[
n-p=\frac{2k_{B}^{2}T^{2}}{\pi\hbar^{2}v_{F}^{2}}\Gamma\left(2\right)\left\{ \mathcal{F}_{1}\left(-\eta_{F}\right)+\mathcal{F}_{1}\left(\eta_{F}\right)\right\} ,\]

which is in agreement with literature \cite{Fang_APL_Q_Cap}. Using
this expression, and differentiating it at the Fermi energy, we obtain

\[
C_{Q}\left(E\right)=\frac{2e^{2}k_{B}T}{\pi\hbar^{2}v_{F}^{2}}\Gamma\left(2\right)\left\{ \mathcal{F}_{0}\left(-\eta_{F}\right)+\mathcal{F}_{0}\left(\eta_{F}\right)\right\} ,\]

which in the low temperature limit, can be shown to be

\[
C_{Q}=\frac{2e^{2}k_{B}T}{\pi\hbar^{2}v_{F}^{2}}\ln\left[2\left(1+\cosh\frac{qV_{TG}}{k_{B}T}\right)\right].\]

This result can be further generalized to a case of a realistic GNR
of finite width, yielding Eq. 3 in the text. 

\bibliographystyle{aipnum4-1}
\bibliography{Graphene_REFS}

\end{document}